# PGA: A Program for Genome Annotation by Comparative Analysis of Maximum Likelihood Phylogenies of Genes and Species

Paulo Bandiera-Paiva[1] and Marcelo R.S. Briones[2]

[1]Departmento de Informática em Saúde and [2]Departamento de Microbiologia, Imunologia e Parasitologia, Universidade Federal de São Paulo, 04023-062, São Paulo, Brazil.

To whom correspondence should be addressed

Paulo Bandiera-Paiva

Ed. Leal Prado, Universidade Federal de São Paulo, Rua Botucatu, 862, Térreo.

CEP 04023-062, São Paulo, S.P., Brazil

e-mail:paiva@unifesp.br

Tel: (55)(11) 5576-4347

FAX: (55)(11) 5571-6504

**ABSTRACT**

The Phylogenetic Genome Annotator (PGA) is a computer program that enables real-time comparison of "gene trees" versus "species trees" obtained from predicted open reading frames of whole genome data. The gene phylogenies are inferred for each individual genome predicted proteins whereas the species phylogenies are inferred from rDNA data. The correlated protein domains, defined by PFAM, are then displayed side-by-side with a phylogeny of the corresponding species. The statistical support of gene clusters (branches) is given by the quartet puzzling method. This analysis readily discriminates paralogs from orthologs, enabling the identification of proteins originated by gene duplications and the prediction of possible functional divergence in groups of similar sequences.

**Availability:** PGA and documentation is freely available under the GNU General Public License and can be downloaded from: http://compbio.epm.br/pga.

**Contact:** pga@compbio.epm.br

**INTRODUCTION**

The characterization of distinct protein functions in a group of sequences sharing a common domain can be improved by comparing the evolutionary history of the underlying species with the inferred phylogeny of the genes (Eisen, 1998). Differences between topologies of these two phylogenies indicate gene duplication events.

The prediction of protein function from primary structure is a common task that can be performed using different methodologies. The most common approach aims to identify similarities between the query sequence and sequences with known function, assuming that similar sequences share a common ancestor and thus include stretches of similar states also present in the precursor sequence. This assumption of orthology based on similarity, disregards the possibility of gene duplication (paralogy) or convergent evolution of the genes. The comparison between gene trees versus species trees is informative about the evolutionary history of genes under study, resulting in a more reliable annotation of predicted proteins of unknown function.

An integrated tool for the analysis of correlated proteins that share a common domain is presented here. This program, PGA, explicitly considers the phylogenies of species and predicted proteins which enables the visualization of gene duplication, or paralogy, and therefore contributing to reveal novel functions in a group of similar sequences.

**DESCRIPTION**

The phylogenetic genome annotator (PGA) uses two coordinated datasets, one of proteins (gene sequence set) and the other of ribosomal RNA from their underlying species (species sequence set). The first dataset is used for inferring the phylogeny of the genes, whereas the second for the inference of the species tree of the source organisms bearing the proteins of the first dataset.

The gene sequence sets are data structures representing groups of related sequences,

containing taxonomic information, primary structure, and a multiple alignment. This structure is used both for protein sequences and to access nucleotide sequences of ribosomal RNA, used for the inference of species phylogeny. The gene sequence set is used for the manipulation of protein sequences representing genes of interest, while the species sequence set maintains the rRNA sequences from the source organisms present in the prior set.

The gene sequence set is specified by the user from a group of related proteins. These sequences may be manually entered, specified by accession id, or automatically fetched from the seed sequences of a specified PFAM protein domain family (Bateman *et al*., 2000). Proteins specified by accession and those pertaining to the PFAM seed sequences (with incomplete information), need to be retrieved by the system from public databases. All necessary sequence data, primary structure and taxonomic information, is retrieved from either the NCBI database using extensible markup language (XML) or from the SWISSPROT database using its data communication interface. All protein sequences included by the user must be related by matching a common protein domain profile or by having significant similarity in their primary structure. Sequences in the gene set are aligned to a hidden Markov model (HMM) profile of a common protein domain. This profile-driven multiple alignment is performed by the hmmalign tool from the HMMER package (Eddy, 2001). Only matching states of the sequences are considered in the alignment. This approach includes only residues which effectively participate in the protein domain being analyzed, thus representing conserved regions. The use of matching states solely in the multiple sequence alignment renders a reliable dataset for the phylogenetic inference since these domains are more likely to be conserved in the evolutionary process.

The underlying species phylogeny is inferred using ribosomal RNA data. The species sequence set contains aligned ribosomal sequences referring to the species parsed from the "source organism" fields in the gene sequence set. The ribosomal RNA sequences of the species list in the gene set are retrieved from the Ribosomal Database Project (RDP-II) (Cole *et* *al.*, 2003). The RDP-II is a repository that contains data from small subunit (SSU) ribosomal RNA of organisms from domains Eukaryota, Bacteria and Archaea. Ribosomal RNA sequences in the RDP-II databases are maintained at the repository in aligned format. These sequences are used for inferring the evolutionary history of the considered organisms and the domain must be specified in the RDP-II database system settings.

The phylogenies of the gene set and the corresponding species set are inferred using the TREE-PUZZLE program package (Schmidt *et al.*, 2002), which uses quartet-puzzling maximum likelihood phylogenetic analysis, including gamma distributed rate heterogeneity. The use of gamma distribution is computationally intensive and can be disabled in PGA if processing time is a limitation. The number of distinct rates and the alpha parameter should also be specified by the user.

The phylogenetic inference of both sets renders polytomic unrooted trees. Gene duplication events are predicted using a methodology previously described (Zmasek and Eddy, 2001), modified to suit polytomic trees. Phylogenies of the gene and species sequence sets are presented in a single window, allowing the visual comparison of both trees. Predicted duplication events are depicted in the gene tree.

**IMPLEMENTATION**

PGA was written in C++ programming language (GNU gcc version 2.96) using Qt, a multi platform graphic user interface (GUI) application framework. Application data, such as sequence sets and the PFAM domain profiles are stored in a relational database based on a MySQL database management system. The tool was developed primarily for use in Intel based Linux workstations and can easily be ported to other platforms.

An example is given in Fig. 1, where an Acethylcholine set is analysed. The main program window displays summary data the sequence set, the number of distinct species and PFAM profile used, in this case 7tm_1 (ACC: PF00001). The sequence set viewer shows the profile-based multiple alignment and taxonomic information and the tree view displays the gene tree on the left and the species tree on the right. The potential gene duplication events (paralogy) are indicated by circles at the internodes. Selection of any sequence in either tree highlights all sequences originated from the same species (light gray in the figure and red in the actual application screen).

Because the PGA maximum likelihood phylogenies are obtained by the quartet-puzzling methods the QP support, or the frequencies of branches in the space of possible quartets is presented in the final, maximum likelihood tree. Accordingly, QP values above 70% indicate potentially supported clusters and frequencies above 90% indicate strongly supported clusters. The QP scores can roughly be compared to Bootstrap values although should not be confused with. We recommend the TREE-PUZZLE documentation for a deeper discussion (Strimmer and von Haeseler, 1996).

## Acknowledgements

P.B.P. received graduate scholarships from UNIFESP (Brazil). This work was supported by grants to M.R.S.B. from FAPESP and CNPq (Brazil), and the International Research Scholars Program of the Howard Hughes Medical Institute (USA).

**Figure legend**

Fig. 1. PGA graphical user interface. The main program window displays summary data of an Acetylcholine sequence set, with the number of distinct species and PFAM profile used. The sequence set viewer shows the profile-based multiple alignment and taxonomic information. The tree view displays the sequence set tree on the left and the species tree on the right. Putative gene duplication events (paralogy) is indicated by circles. Selection of any sequence in the trees highlights all sequences originated from the same species (light gray in the figure and red in the actual application screen). Numbers above branch clusters indicate the quartet puzzling (QP support), where values above 70% indicate potentially supported clusters and frequencies above 90% indicate strongly supported clusters. The QP scores can roughly be compared to Bootstrap values although should not be confused with, please see TREE-PUZZLE documentation for a deeper discussion.

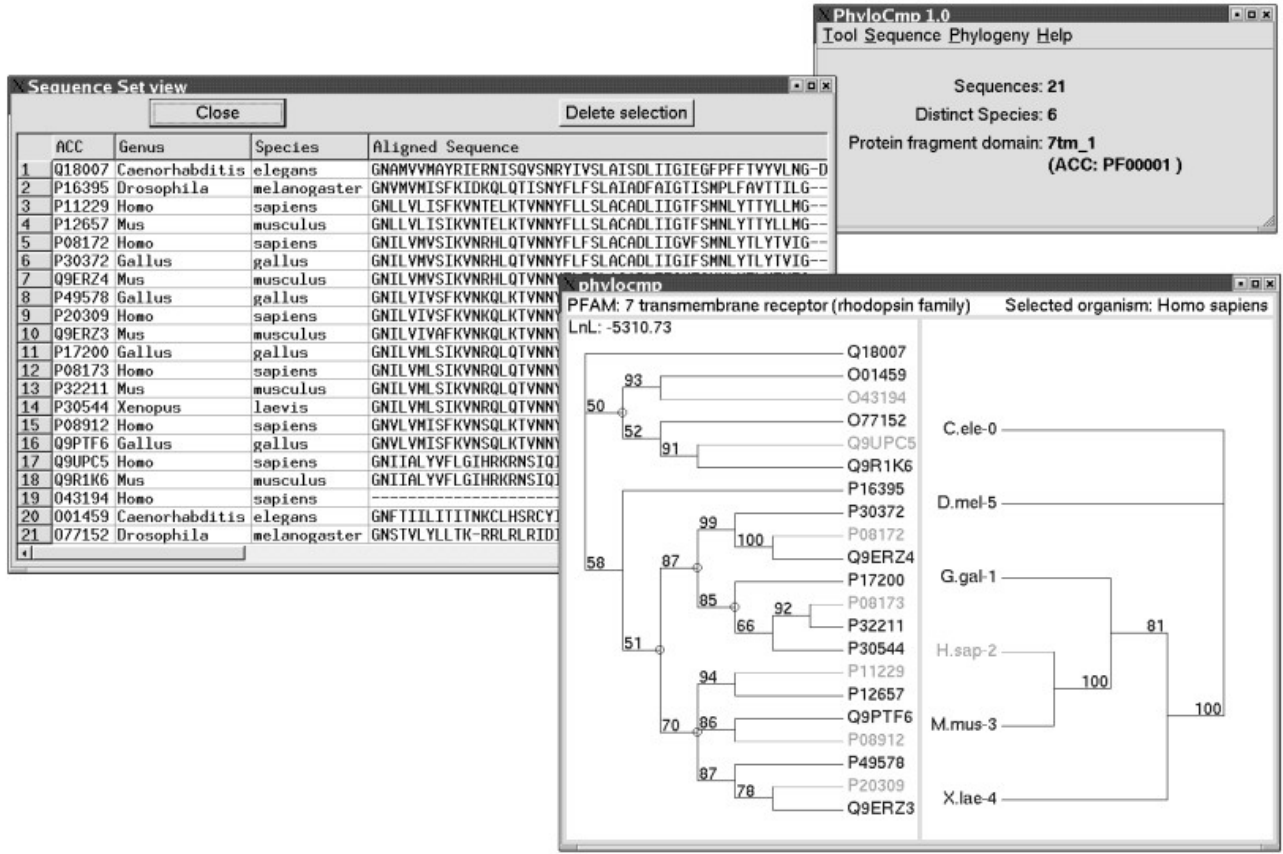

Fig. 1.